\documentclass[aip,graphicx,reprint,longbibliography,noeprint]{revtex4-1}

\usepackage[pdftex]{graphics}
\usepackage{amssymb,amsmath}
\usepackage{bm}
\usepackage{siunitx}

\newcommand{\rd}{\mathrm{d}}
\newcommand{\ii}{\mathrm{i}}
\newcommand{\e}{\mathrm{e}}
\newcommand{\Tr}{\mathrm{Tr}}
\newcommand{\rev}[1]{\textcolor{black}
{#1}} 
\newcommand{\revb}[1]{\textcolor{black}
{#1}} 
\renewcommand{\Re}{\operatorname{Re}}

\begin{document}

\title{A multi-state mapping approach to surface hopping}
\author{Johan E. Runeson}
\email{johan.runeson@chem.ox.ac.uk}
\affiliation{Department of Chemistry, University of Oxford, Physical and Theoretical Chemistry Laboratory, South Parks Road, Oxford, OX1 3QZ, UK}
\author{David E. Manolopoulos}
%\email{david.manolopoulos@chem.ox.ac.uk}
\affiliation{Department of Chemistry, University of Oxford, Physical and Theoretical Chemistry Laboratory, South Parks Road, Oxford, OX1 3QZ, UK}

\begin{abstract}
We describe a multiple electronic state adaptation of the mapping approach to surface hopping introduced recently by Mannouch and Richardson (J. Chem. Phys. 158, 104111 (2023)). Our modification treats populations and coherences on an equal footing and is guaranteed to give populations in any electronic basis that tend to the correct quantum--classical equilibrium values in the long-time limit (assuming ergodicity). We demonstrate its accuracy by comparison with exact benchmark results for three- and seven-state models of the Fenna–Matthews–Olson complex, obtaining electronic populations and coherences that 
are significantly more accurate than those of fewest switches surface hopping and at least as good as those of any other semiclassical method we are aware of. \rev{Since these results were obtained by adapting the scheme of Mannouch and Richardson, we go on to compare our results with theirs for a variety of problems with two electronic states. We find that their method is sometimes more accurate, and especially so in the Marcus inverted regime. However, in other situations the accuracies are comparable, and since our scheme can be used with multiple electronic states it can be applied to a wider variety of electronically nonadiabatic systems.}
\end{abstract}

\maketitle

\section{Introduction}

Fewest switches surface hopping (FSSH) is the dominant method for simulating electronically nonadiabatic dynamics in chemistry. Originally proposed by Tully in the early 1990's,\cite{tully1990hopping} it has become a standard tool for running {\em ab initio} trajectories and is now routinely used to study various aspects of photochemical dynamics including both energy and charge transfer processes.\cite{Barbatti2018review,Menger2018exciton,Peng2022exciton,Toldo20S23fssh_review}
It remains popular for three (good) reasons: (i) it is easy to use; (ii) it is robust when used with {\em ab initio} potentials; and (iii) it has been found to work reasonably well in many applications. However, while there have been attempts to derive it from first principles,\cite{subotnik2013QCLE} FSSH is still widely considered to be an \emph{ad hoc} algorithm. It also suffers from a well-known `overcoherence' problem that has led to the development of a number of different `decoherence' corrections,\cite{wang2020review} none of which has become universally accepted.

A second school of thought advocates treating the electronic and nuclear degrees of freedom on an equal footing by mappling the multi-state electronic system onto a classical phase space. The simplest approach of this type is (multiple-trajectory) Ehrenfest dynamics, which is based on a mean-field coupling of the nuclear and electronic motions. However, this is known to produce various unphysical results including an overheating of the electronic subsystem.\cite{parandekar2006ehrenfest} Semiclassical mapping methods represent the multiple electronic states as harmonic oscillators\cite{meyer1979nonadiabatic,stock1997mapping} or generalized spins,\cite{runeson2020} and use either initial phase space sampling or symmetrical quasi-classical binning\cite{cotton2013a,miller2016review} to calculate the dynamics. These mapping methods typically give better accuracy than either FSSH or Ehrenfest dynamics in applications to system--bath models of condensed-phase systems. However, they do so at the cost of allowing the electronic-state populations to become negative, which can result in the nuclei evolving (and potentially even diverging) on inverted potentials. While this problem may not be visible in simple models, it does help to explain why mapping methods are only rarely combined with {\em ab initio} potentials and applied to realistic chemical systems.

Given the complementary strengths and weaknesses of surface hopping and phase-space mapping, it is natural to ask whether they could be combined to give a better method. Very recently, Mannouch and Richardson have explored this idea and used it to develop a new `mapping approach to surface hopping' (MASH).\cite{Mannouch2023mash} As in surface hopping, this method uses trajectories that hop between the physical adiabats, thereby eliminating concerns about dynamics on inverted mapped potentials. But rather than employing stochastic hops, the active surface is obtained deterministically from the electronic wavefunction, as it is in the phase-space mapping approach. This has several distinct advantages, not least of which is that it avoids the need for heuristic `decoherence' corrections. The results can instead be improved (if necessary) through a careful resampling of the electronic wavefunction (the quantum-jump procedure\cite{kapral2015quantum}).
Even without this resampling, MASH has been found to give better results than either pure surface hopping or pure mapping for a wide range of two-state problems, with no more computational effort.\cite{Mannouch2023mash} It is thus arguably the most promising method that has yet been proposed for nonadiabatic dynamics.

However, the formulation of MASH that Mannouch and Richardson presented was restricted to two coupled electronic states.\cite{Mannouch2023mash} It is not obvious how to extend it to more states because they based their formulation on correlation functions involving specific two-state prescriptions for the electronic populations and coherences that were specialized to the adiabatic basis (see Sec~II). A general nonadiabatic dynamics method should be applicable to an arbitrary number of electronic states, it should treat populations and coherences on an equal footing, it should provide results that transform correctly under the unitary rotations that take one electronic basis to another, and it should be able to provide these results directly in any chosen basis. In Sec.~III, we describe such a multi-state adaptation of MASH, and show that it is guaranteed to give the correct quantum--classical equilibrium populations of the electronic states in any basis in the long-time limit (assuming ergodicity). In Sec.~IV, we demonstrate that this adaptation works just as well for a standard multi-state exciton energy transfer problem as Mannouch and Richardson have shown MASH to work for two-state problems.\cite{Mannouch2023mash} \rev{In Sec.~V we compare our adaptation with Mannouch and Richardson's original version of MASH for a variety of problems with just two electronic states, and in Sec.~VI we conclude this paper.}

%However, their method is so far only applicable to two-level systems. It is not obvious how to generalize it to multiple levels, partly because the method in its current formulation relies on a rather complicated construction of time-dependent correlation functions. Specifically, Mannouch and Richardson measured populations and coherences on a different footing and all observables have to be first computed in the adiabatic basis and then transformed to the basis of interest.

%The present article makes two contributions to overcome these limitations. Firstly, we show that MASH can be applied to general $N$-state systems. Secondly, we simplify calculations by introducing a new estimator which (i) does not discriminate between populations and coherences, (ii) can therefore be directly applied in any basis, and (iii) guarantees the correct long-time equilibrium in any basis. We apply the method to exciton transfer in the Fenna--Matthews--Olson complex and find that it reproduces fully quantum calculations more accurately than any other method that we are aware of.

\section{Nonadiabatic dynamics}

Consider a general nonadiabatic system defined by the Hamiltonian
\begin{equation}
\hat{H}(p,q) = \sum_{j=1}^f \frac{p_j^2}{2m_j} + \hat{V}(q),
\end{equation}
where $q=\{q_j\}$ and $p=\{p_j\}$ are the coordinates and momenta of $f$ nuclear degrees of freedom with masses $\{m_j\}$, and $\hat{V}(q)$ is the potential energy operator. In model systems, it is usually most convenient to express the potential in a given diabatic ($q$-independent) basis $\{|n\rangle\}$, i.e.,
\begin{equation}
\hat{V}(q) = \sum_{nm} V_{nm}(q) |n\rangle\langle m|.
\end{equation}
%Although the diabatic basis is the most convenient starting point for model systems, it is usually not known \emph{a priori} in \emph{ab initio} applications.
Although it would in principle be possible to evolve surface hopping dynamics directly in the diabatic basis, it is in practice almost always run in the adiabatic basis (the local eigenbasis of $\hat{V}(q)$), where
\begin{equation}
\hat{V}(q) = \sum_a V_a(q) |a(q)\rangle\langle a(q)|.
\end{equation}
The two bases are related through their transformation matrix elements $U_{na}(q)=\langle n|a(q)\rangle$. % latter is related to the nonadiabatic coupling vector $d_{ab}^i\equiv \langle a(q)|\nabla_j|b(q)\rangle$ through 
In this article, we will consider model systems defined in a diabatic basis, but our method can also be used directly with potentials in the adiabatic basis, such as those provided by \emph{ab initio} electronic structure calculations.

For most photochemical applications, it is a reasonable approximation to treat the nuclei as classical particles. % (in any case, treating them fully quantum-mechanically would be unfeasible for anything but simple model systems).
%(A full quantum-mechanical treatment is only possible for simple model systems) 
However, to consistently evolve a coupled system of classical and quantum degrees of freedom is a long-standing problem in semiclassical dynamics. To see why, consider a particle at configuration $q$ with momentum $p$ and with electronic state $|\psi\rangle=\sum_n c_n |n\rangle$. The natural starting point is to evolve the electronic state according to Schr\"{o}dinger's equation of motion. This can be done equivalently either in the diabatic basis
\begin{equation}\label{eq:dotc_dia}
\dot{c}_n = -\ii \sum_m V_{nm}(q)c_m,
\end{equation}
or in the adiabatic basis
\begin{equation}
\dot{c}_a = -\ii V_a(q) c_a - \sum_j \frac{p_j}{m_j}\sum_b d_{ab}^j(q)c_b,
\end{equation}
where $d^j_{ab}(q) = \langle a(q)|\nabla_j|b(q)\rangle$ is an element of the nonadiabatic coupling vector. (Throughout this paper we use units where $\hbar=1$.)
The main difficulty arises when constructing the nuclear dynamics, and in particular when considering the `back-action' of the electrons on the nuclei. There is a `force operator' $\hat{F}_j(q)=-\nabla_j \hat{V}(q)$, but this is not yet useful to run classical dynamics. What one would like is equations of motion of the form
\begin{subequations}
\begin{align}
\dot{q}_j &= p_j/m_j \\
\dot{p}_j &= F_j(q),
\end{align}
\end{subequations}
where $F_j(q)$ is a (yet to be defined) classical force. 
Existing schemes for nonadiabatic dynamics differ mainly in the way they construct this force. 
In the following subsections we briefly summarize and comment on some of the most important strategies.

\subsection{Fewest switches surface hopping}\label{sec:fssh}
%Tully's fewest switches surface hopping is the dominant method for nonadiabatic dynamics in photochemistry. Although there have been attempts to derive it as an approximation to more accurate theories, it is still broadly considered an \emph{ad hoc} method, which remains popular because it is (i) easy to apply and (ii) has been found to work decently well for many problems. 

In surface hopping, the instantaneous force on the nuclei is taken to be that of a single adiabatic state, called the \emph{active surface}. If the active surface is $a$, then this force is
\begin{equation}\label{eq:FMASH}
F_j(q) = -\langle a(q)|\nabla_j \hat{V}(q)| a(q)\rangle.
\end{equation}
If the trajectories enter a region with non-zero nonadiabatic coupling, they can switch active surface (or `hop'). At each discrete time step, the probability to hop from $a$ to $b$ is taken to be
\begin{equation}
P_{a\to b} = 2\Delta t \Re\left(\frac{c_b}{c_a}\right) \sum_j \frac{p_j}{m_j} d_{ab}^j.
\end{equation}
Tully chose this probability such that the fraction of trajectories evolving on surface $a$ would approximate the average population $\langle |c_a|^2\rangle $ with a minimal number of switches.\cite{tully1990hopping} If a hop occurs, the momentum is rescaled along the nonadiabatic coupling vector such that the total energy is conserved. If there is not sufficient kinetic energy to overcome the difference in potential energy between the pre- and post-hop adiabatic surfaces, the standard (although not universally accepted \cite{martens2016CSH}) practice is to reject the hop and reverse the momentum in the direction of the nonadiabatic coupling vector.

Despite the considerations behind the choice of hopping probability, the fraction of trajectories evolving on surface $a$ does not strictly agree with $\langle |c_a|^2\rangle$. This inconsistency is at the root of many of the issues present in surface hopping.\cite{subotnik2016review,Carof2017FOB-SH} Traditionally, the problem has been identified as an `overcoherence' of the electronic coefficients that can be overcome with (more or less heuristic) `decoherence corrections'. Many such corrections have been proposed, but despite much effort, there is as yet no consensus as to whether any of them has solved the underlying problem. Furthermore, surface hopping does not generally guarantee relaxation to the correct long-time equilibrium in condensed-phase systems, although it does often provide a better approximation than Ehrenfest dynamics.\cite{parandekar2005mixed,schmidt2008SH}

%They may improve accuracy for certain cases, but can also ruin the trajectories if applied
%[Not always adiabatic: \cite{Truhlar2023gradcorr}]

\subsection{Semiclassical mapping approaches}

A rather different strategy is to construct the force to be a coherent average over contributions from multiple electronic states. A simple approach of this type is Ehrenfest dynamics, which uses the expectation value of the force operator,
\begin{equation}\label{eq:ehrenfest}
F_j(q) = -\langle \psi|\nabla_j \hat{V}(q)|\psi\rangle = -\sum_{nm} \nabla_j V_{nm}(q) c_n^* c_m.
\end{equation}
Ehrenfest dynamics has several severe drawbacks, of which the most important are that it violates detailed balance (it relaxes the system to an overheated equilibrium) and fails to capture wavepacket branching in scattering models.
Nevertheless, it has the advantage of being invariant to a unitary transformation of the electronic basis,
%(whereas surface hopping requires a particular choice of basis) 
and the deterministic nature of the force allows an ergodic analysis of its long-time limit. Explicitly, the real and imaginary parts of the electronic coefficients can be regarded as phase-space variables on the same footing as the nuclear degrees of freedom. In terms of these variables, it is clear that Eq.~\eqref{eq:dotc_dia} is equivalent to the dynamics of a set of $N$ harmonic oscillators.

The last of these observations has inspired a more formal mapping of the electronic states to quantum harmonic oscillators, which is now known as the Meyer--Miller--Stock--Thoss (MMST) mapping. \cite{meyer1979nonadiabatic,stock1997mapping} Taking the classical limit of this oscillator model leads to a classical phase-space theory with its own force as well as expressions for population and coherence estimators. The new force is also of coherent-average type, but differs from the Ehrenfest force in that the instantaneous populations can be negative (or larger than one). Despite this seemingly unphysical behaviour, the weighted average over many trajectories has for many model problems been found to be more accurate with the MMST mapping than in (multi-trajectory) Ehrenfest dynamics.\cite{stock2005nonadiabatic,miller2016review} 

Harmonic oscillators are not the only way to map the electronic coefficients onto classical variables. For two-level systems, another choice would be to use the well-known isomorphism to a spin-1/2 system.
%\begin{equation}
%\hat{V}(q) = H_0(q) \hat{\sigma}_0 + \sum_{i=1}^3 H_i(q) \hat{\sigma}_i.
%\end{equation}
This approach was recently used to develop a `spin mapping' analogous to the MMST mapping,\cite{runeson2019} which leads to a subtly different definition of the force and the estimators. For $N$-level systems, the spin mapping has a natural generalization in terms of the so-called Stratonovich--Weyl transformation.\cite{runeson2020} This has been shown to at least partly solve the overheating problem of Ehrenfest dynamics, in the sense that the long-time equilibrium reduces to phase-space averages that agree with quantum mechanics up to first order in $\beta=1/k_{\rm B}T$.\cite{runeson2022chimia} 
%This property is unique to spin mapping and is not present in any of the approaches based on MMST mapping.\cite{amati2023}
%In model problems, the spin-mapping version has been found to be generally more accurate than its MMST counterpart.

However, just like the MMST mapping, spin mapping can predict (unphysical) negative populations at low temperatures. For an individual trajectory, an instantaneous negative population can cause it to evolve and diverge on an inverted potential, which is completely unphysical. A recent attempt to overcome this problem was to construct a mapping to an anisotropic spin.\cite{Amati2023ellipsoid} This does remove the issue of inverted potentials and can (at least for two-level systems) be constructed so as to give the correct long-time equilibrium populations at any temperature, as long as the classical nuclear assumption is valid. However, the timescale of the relaxation to equilibrium was  found to be worse in several cases than that of the original spin mapping.

%Hence, surface hopping and mapping approaches each have their own advantages and disadvantages. This raises the question whether it is possible to construct a method that combines the strengths of both strategies. A recent article in \emph{J. Chem. Phys.} suggests that the answer could be yes.

\subsection{Mapping approach to surface hopping} 

It should be clear from what we have said so far that the surface hopping and mapping approaches each have their own advantages and disadvantages. Given this, the natural question is whether it is possible to develop a method that combines the strengths and eliminates the weaknesses of the two strategies.

Mannouch and Richardson have recently shown that, at least for two-level systems, this may indeed be possible.\cite{Mannouch2023mash} Their  `mapping approach to surface hopping' (MASH) uses the nuclear force of a single active adiabatic surface as in surface hopping,
but instead of treating the active surface $V_a(q)$ separately from the electronic wavefunction, they set it to be that of the adiabatic state with the \emph{largest instantaneous population} $|c_a|^2$. In this way, there is no need to introduce a stochastic hopping probability, because the active state is always uniquely determined by the electronic wavefunction. %Moreover, the inconsistency between the active surface $a$ and the stochastically averaged probability $\langle |c_a|^2\rangle$ in traditional surface hopping is eliminated entirely. 

\begin{figure}[t]
\centering
\resizebox{0.65\columnwidth}{!} {\includegraphics{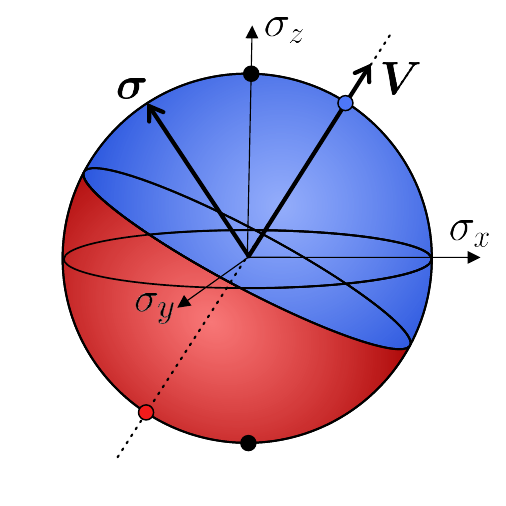}}
\caption{Bloch sphere representation of a two-level system. Any selection of two opposite fixed points on the sphere (e.g., the black dots) defines a diabatic basis, whereas the adiabatic basis corresponds to the opposite points on an axis that follows $(V_x,V_y,V_z)$ (blue and red dots). In MASH, the nuclei evolve on the adiabatic state that is closest to the instantaneous direction of the Bloch vector, $\bm{\sigma}=\langle\psi |\hat{\bm{\sigma}}|\psi\rangle$.}\label{fig:bloch}
\end{figure}

Figure~\ref{fig:bloch} illustrates the situation on the Bloch sphere of a two-level system. Adiabatic states 1 and 2 correspond to the opposite poles along the axis parallel to $\bm{V}=(V_x,V_y,V_z)$, where $V_\alpha=\Tr[\hat{V}\hat{\sigma}_\alpha]$ and the $\{\hat{\sigma}_\alpha\}$ are the Pauli spin operators. In MASH, the instantaneous active surface is set to 1 when the Bloch vector $\bm{\sigma}=\langle\psi |\hat{\bm{\sigma}}|\psi\rangle$
is on the hemisphere closest to adiabatic state 1, and 2 when it is on the hemisphere closest to adiabatic state 2.

In addition to \rev{providing a deterministic alternative to stochastic surface hopping}, the MASH approach has several other appealing features, a more detailed discussion of which can be found in Ref.~\onlinecite{Mannouch2023mash}: 
\begin{enumerate}
\item[(i)] it satisfies detailed balance\cite{Amati2023ellipsoid} in the sense that it gives the correct long-time populations of the adiabatic electronic states (assuming ergodicity);
\item[(ii)] the momentum rescaling and momentum reversal arise naturally in the deterministic surface hopping algorithm without any ambiguity;
\item[(iii)] in place of heuristic decoherence corrections, there is a way to systematically improve the results by carefully resampling the electronic wavefunction along the trajectory (the quantum jump prodecure).
\end{enumerate}

Mannouch and Richardson have demonstrated by comparison with quantum benchmark calculations that MASH is more accurate for typical system-bath models than both FSSH and state-of-the-art mapping approaches.
%[If Joe's paper is out, mention rates] 
Their method can treat wavepacket branching just at least as well as FSSH, which is currently only possible in mapping via a more expensive cancellation of positive and negative phase-space contributions.\cite{runeson2021} They also found MASH to be the most accurate classical-trajectory method considered so far in describing ultrafast internal conversion in pyrazine.\cite{Mannouch2023mash}

However, their formulation of the method was restricted to two-level systems,\cite{Mannouch2023mash} and it is not obvious how to generalize it to more electronic states because Mannouch and Richardson specifically constructed their observables for the case of two. Explicitly, for electronic observables $\hat{A}$ and $\hat{B}$, they computed the time-correlation function 
\footnote{In their paper, Mannouch and Richardson used spin vectors rather than $c$ coordinates, but these contain the same information and are simply a different a notation for the same quantity.}
\begin{equation}\label{eq:CWAB}
C_{AB}(t) = \langle \mathcal{W}_{AB}(c_0) A(c_0) B(c_t) \rangle, 
\end{equation}
where $\langle \cdot \rangle$ is an expectation value taken with respect to an appropriate density function over $(p_0,q_0,c_0)$. Here the estimators $A(c)$ and $B(c)$ are constructed from the following complete set of mappings for two-level populations and coherences defined in the adiabatic representation, 
\begin{subequations} \label{eq:mash_observables}
\begin{align}
|1\rangle\langle 1| &\mapsto h(|c_1|^2-|c_2|^2) \\ 
|2\rangle\langle 2| &\mapsto h(|c_2|^2-|c_1|^2)\\
|a\rangle \langle b| &\mapsto c_a^* c_b  \quad\ \, (a\neq b), 
\end{align}
\end{subequations}
where $h(x)$ is the Heaviside step function. The weight function $\mathcal{W}_{AB}(c)$ is also specified in the adiabatic representation, as $\mathcal{W}_{AB}(c) = 3$ if $A$ and $B$ are both coherences, 2 if one is a coherence and the other a population, and $2|(|c_1|^2-|c_2|^2)|$ if they are both populations. Diabatic observables first have to be converted to the adiabatic representation, as in conventional surface hopping, in order to calculate these quantities.

Although this scheme has been shown to work well for a multitude of problems, it is worth asking whether there exists a simpler procedure, especially if one aims to generalize it to more than two electronic states. Firstly, in our view, there should be no reason to give coherences a special treatment. Since these are just population differences in a rotated basis, one should be able to treat them on an equal footing to the populations. Secondly, it would be preferable to calculate the observables in any diabatic basis directly from the wavefunction coefficients in that basis, without having to first convert them to a linear combination of adiabatic observables. The adaptation of MASH presented in the next section achieves both of these goals while at the same time generalising the method to an arbitrary number of coupled electronic states.

\section{Multi-state mapping} %approach to surface hopping}

Having set the context of what we are about to do, we are now ready to propose a multi-state generalisation of MASH. We will separately address the issues of how to run the dynamics, how to measure observables, and how to set up the initial conditions for a typical photochemically initiated nonadiabatic problem. 

\subsection{Dynamics}

The dynamics extends quite naturally from the two-level case by using the same expression for the nuclear force as in Eq.~\eqref{eq:FMASH} and picking the active surface to be that of the adiabatic state with the largest instantaneous population. 
To express this more precisely, we define the $N$ populations and phases in a given basis as
\begin{subequations}
\begin{align}
P_n & =|c_n|^2 \\
\phi_n &=\arg c_n. 
\end{align}
\end{subequations}
If we also define the classical state projectors 
\begin{equation}\label{eq:Thetadef}
\Theta_n = \begin{cases} 
1 & \text{if } P_n > P_m ~\forall\, m\neq n \\
0 & \text{otherwise}, 
\end{cases}
\end{equation}
then precisely one of these $N$ projectors will be non-zero in any basis at each point in the classical $c$ space. As in two-level MASH, we take the dynamically active potential energy surface -- the surface that is used to define the nuclear force in Eq.~\eqref{eq:FMASH} -- to be that of the unique adiabatic basis state $a$ with a non-zero classical state projector $\Theta_a$. This choice of the adiabatic potential for the nuclear dynamics is consistent with experience from decades of successful surface-hopping calculations. 

\begin{figure}[t]
\centering
\resizebox{0.9\columnwidth}{!} {\includegraphics{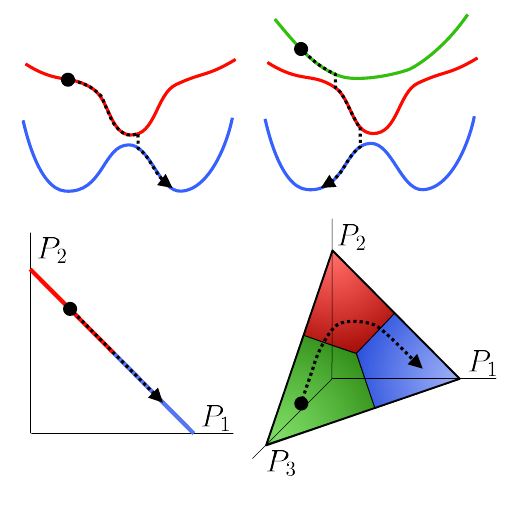}}
\caption{Schematic depiction of a MASH trajectory for two (left) and three (right) states. In any given basis, the set of $N$ populations is a point on a simplex (generalized triangle) with the $N$ unit vectors as vertices. In MASH, the active surface is set to be the adiabatic state with the highest instantaneous population. Thereby, each adiabatic state corresponds to the region on the simplex closest to a given vertex. Hops occur whenever the system crosses a border between two of these regions.}\label{fig:simplex}
\end{figure}

Assuming the initial electronic wavefunction is normalized, the populations will always sum up to one in any basis, $\sum_{n=1}^N P_n = \sum_{n=1}^N|c_n|^2=1$, and they will always be non-negative, $P_n\geq 0$. (The electronic evolution in Eq.~\eqref{eq:dotc_dia} is unitary and preserves these two conditions for all time.)
In the two-level case, one can visualize $(P_1,P_2)$ as a point evolving on the line segment between $(1,0)$ and $(0,1)$. Similarly, in the three-level case $(P_1,P_2,P_3)$ is a point on the triangle with vertices $(1,0,0)$, $(0,1,0)$, and $(0,0,1)$. Figure~\ref{fig:simplex} illustrates the regions of different active surfaces for these two cases. 
In general, for $N$ levels, $(P_1,P_2,\dots,P_N)$ is a point on the \emph{simplex} 
\begin{equation}\label{eq:Simplex}
 \{ P \in \mathbb{R}^N | \sum_n P_n = 1, P_n \geq 0 ~\forall\, n \}
\end{equation}
which is a geometrical object that can be thought of as a high-dimensional triangle with vertices on the unit vectors.\cite{bengtsson2017geometry}

When a trajectory passes from the region associated with adiabatic state $a$ to that associated with state $b$, there is a `hop', and we rescale the momentum along a \rev{direction determined by the nonadiabatic coupling vector as explained in detail in Appendix~E. 
If the kinetic energy is insufficient to hop, we reverse the momentum along the same direction. So far, everything is just a direct extension of} %identical to 
the two-level case.\cite{Mannouch2023mash} In the three-level case, however, we need to address a new issue, which is that hops can occur between states that are uncoupled from each other. For example, if states 1 and 2 are coupled but both are uncoupled from state 3, then trajectories that start in the blue region but are close enough to the green can hop to state 3 before they reach state 2. It is not obvious how to deal with this situation. The simplest option is to accept it. Another option would be to reject the hop if the coupling between the two relevant states (in this case 1 and 3) is less than a given threshold. As in the case of insufficient kinetic energy, one would in such a situation also reverse the momentum. 
%An interesting, although not recommended, possibility would be to make the trajectory remember its history and keep moving on state 1 even though it is in region 3, until it hits a region of sufficient coupling between the states. However, this approach would most certainly break detailed balance and will not be considered in the following.
Here we will adopt the simplest option and accept the hop regardless of how strongly coupled the states are. This avoids the ambiguity of defining a coupling threshold and it finishes our discussion of the dynamics. % and leaves the dynamics free of additional parameters.
What remains to be decided is how to measure observables. Since our aim is a method that gives the correct quantum-classical equilibrium populations in the long-time limit, we shall begin by discussing what happens when the dynamics has reached equilibrium.

%For any $\bm{P}$, the state functions sum up to unity in any basis, $\sum_{n=1}^N \Theta_n(\bm{P})=1$.\footnote{The points where all populations are equal form a set of measure zero and need not be considered.} 

\subsection{Equilibrium}\label{sec:eql}
Because MASH is deterministic (in contrast to standard surface hopping), one can make more powerful statements about its long-time limit. Provided the system is sufficiently ergodic (as is typically true in the condensed phase), any initial distribution will in the long-time limit reduce to the equilibrium distribution 
%Regarding detailed balance, one can for ergodic systems predict the long-time limit of all observables, because they will tend to the distribution
\begin{equation}
\rho(p,q,c) \propto \e^{-\beta E(p,q,c)},
\end{equation}
where the energy function is
\begin{equation}
E(p,q,c) = T(p) + \sum_a V_a(q) \Theta_a
\end{equation}
with $T(p)=\sum_j p_j^2/2m_j$ and $\Theta_a\equiv \Theta_a(c)$. 
%(For brevity, we will in the following 
This means that one can predict the long-time values of all observables in terms of phase-space averages with respect to this distribution. \rev{Such an analysis has recently been done in the two-state case for a variety of mapping methods, including MASH.\cite{Amati2023thermalization} In the following, we consider equilibrium expectation values in MASH for a general $N$-level system.} 

For a coupled electronic--nuclear system, the quantum-mechanical partition function with classical nuclei is
\begin{equation}
Z = \int \frac{\rd p \rd q}{(2\pi)^f} \, \Tr[\e^{-\beta\hat{H}(p,q)}].
\end{equation}
This expression mixes a classical phase-space integral with a quantum trace. MASH (like other mapping approaches) effectively replaces the trace with an integral over all normalized electronic wavefunctions $c$,
\begin{equation}
Z_{\rm MASH} = \int \frac{\rd p \rd q}{(2\pi)^f} \int_{|c|=1} \frac{\rd c}{\mathcal{N}} \, \e^{-\beta E(p,q,c)},
\end{equation}
where $\mathcal{N}=2\pi^N/N!$ normalizes the electronic integral such that $\int_{|c|=1} {\rd c}/{\mathcal{N}}=N$ (see Appendix~A). Because each $\Theta_a$ is an idempotent projector onto a different region of $c$ space, and it has a unit integral $\int_{|c|=1}{\rd c}/{\mathcal{N}}\,\Theta_a=1$, it follows that 
\begin{equation}
\int_{|c|=1} \frac{\rd c}{\mathcal{N}} \, \e^{-\beta \sum_a V_a(q)\Theta_a} =
\sum_a  \e^{-\beta V_a(q)},
\end{equation}
%\begin{multline}
%\int_{|c|=1} \frac{\rd c}{\mathcal{N}} \, \e^{-\beta \sum_a V_a\Theta_a} \\= %\sum_a \int_{|c|=1} \frac{\rd c}{\mathcal{N}}  \, \e^{-\beta V_a}\Theta_a =
%\sum_a  \e^{-\beta V_a}.
%\end{multline}
which shows that MASH is consistent with the mixed quantum-classical partition function ($Z_{\rm MASH}=Z$).
%where the trace is taken over the electronic degrees of freedom. %For each local configuration, $\Tr[\e^{-\beta \hat{H}}]=\sum_a \e^{-\beta E_a}$, where $E_a=T+V_a$. 
By the same argument, the thermal equilibrium population of each adiabatic quantum state,
\begin{equation}
\langle {P}_a\rangle = \frac{1}{Z} \int \frac{\rd p \rd q}{(2\pi)^f} \,\Tr[\e^{-\beta \hat{H}(p,q)}|a(q)\rangle\langle a(q)|],
\end{equation}
will be equal to the corresponding MASH expression
\begin{equation}
\langle \Theta_a\rangle = \frac{1}{Z} \int \frac{\rd p \rd q}{(2\pi)^f} \int_{|c|=1} \frac{\rd c}{\mathcal{N}} \, \e^{-\beta E(p,q,c)} \Theta_a,
\end{equation}
provided $\Theta_a$ is a state projection in the adiabatic representation.
%is equal to $\langle \hat{P}_a\rangle$ provided $\Theta_a$ is a state projection in the adiabatic representation. 
Hence, if one were to choose to use the state projectors as population estimators, this would be guaranteed to give the correct equilibrium populations in the adiabatic basis.

However, it would not work more generally, because $\langle \Theta_n\rangle$ does not equal $\langle P_n\rangle$ in other bases. To see this, we can write the latter explicitly as
\begin{align}\label{eq:EPn}
\langle P_n\rangle &=  \frac{1}{Z} \int \frac{\rd p \rd q}{(2\pi)^f} \,\Tr[\e^{-\beta \hat{H}(p,q)}|n\rangle\langle n|] \nonumber\\
&= \frac{1}{Z} \int \frac{\rd p \rd q}{(2\pi)^f} \, \sum_a \e^{-\beta [T(p)+V_a(q)]} |\langle a(q)|n\rangle|^2, 
\end{align}
whereas the former is
\begin{align}\label{eq:EThetan}
\langle \Theta_n\rangle &=  \frac{1}{Z} \int \frac{\rd p \rd q}{(2\pi)^{f}} \int_{|c|=1} \frac{\rd c}{\mathcal{N}} \, \e^{-\beta E(p,q,c)} \Theta_n \nonumber\\
&=  \frac{1}{Z} \int \frac{\rd p \rd q}{(2\pi)^{f}} \sum_a \e^{-\beta [T(p)+V_a(q)]}\int_{|c|=1} \frac{\rd c}{\mathcal{N}} \,  \Theta_a \Theta_n .
\end{align}
This is in general \emph{not} equal to $\langle P_n\rangle$ because 
\begin{equation} \label{eq:Thetainequality}
\int \frac{\rd c}{\mathcal{N}} \, \Theta_a\Theta_n\neq |\langle a(q)|n\rangle|^2. 
\end{equation}
In other words, the state projection defined by Eq.~\eqref{eq:Thetadef} does not transform correctly as a population estimator under unitary basis rotations.

\subsection{Populations}

A population estimator $\Phi_n$ can in fact be constructed such that $\langle\Phi_n\rangle=\langle P_n\rangle$ in {\em any} basis of $N$ orthonormal electronic states, whether they are adiabatic or diabatic. Here we shall give an overview of the argument that establishes this and leave the technical details to Appendices~B and~C.

Since the Ehrenfest populations $P_n=|c_n|^2$ {\em do} transform correctly (i.e., quantum mechanically) under unitary basis rotations, as do scalars, our ansatz is an equivariant population estimator of the form 
\begin{equation}\label{eq:Phimap}
|n\rangle\langle n| \mapsto \Phi_n = \alpha_N\,P_n+\beta_N,
\end{equation}
in which $\alpha_N$ and $\beta_N$ are scalar parameters that are to yet be determined. Our goal is to choose these parameters such that
\begin{equation}\label{eq:constraint1}
    \int_{|c|=1} {\rd c\over\mathcal{N}}\,\Theta_a\Phi_n = |\langle a(q)|n\rangle |^2
\end{equation}
holds for all adiabatic states $a$ and diabatic states $n$. This will suffice to ensure -- by virtue of the argument in Eqs.~\eqref{eq:EPn} and~\eqref{eq:EThetan} -- that the expectation value of $\Phi_n$  will be correct at equilibrium. Appendix B shows that Eq.~\eqref{eq:constraint1} will be satisfied if the simpler condition
\begin{equation}\label{eq:constraint2}
\int_{|c|=1} \frac{\rd c}{\mathcal{N}} \, \Theta_a\Phi_b = \delta_{ab}
\end{equation}
holds for all adiabatic states $a$ and $b$. The constraints that this imposes on $\alpha_N$ and $\beta_N$ are then investigated in Appendix~C, which obtains the unique solution
\begin{equation}\label{eq:alphadef}
    \alpha_N={N-1\over H_N-1} \quad\hbox{and}\quad \beta_N={1-\alpha_N\over N},
\end{equation}
where $H_N=\sum_{n=1}^N 1/n$. Hence we arrive at the estimator
\begin{equation}\label{eq:Phidef}
\Phi_n = {1\over N}+\alpha_N\left(P_n-{1\over N}\right),
\end{equation}
which has the pleasingly democratic interpretation of  measuring the population of state $n$ relative to the situation in which all the populations are equal to $1/N$ (as they are for the states at the centre of the simplex). 

\subsection{Coherences}\label{sec:coherences}

In contrast to Ref.~\onlinecite{Mannouch2023mash}, as well as most other literature on surface hopping, we shall begin by rewriting coherences as population differences between rotated states so that they can be treated in the same way as populations. Note that we are in an ideal position to do this here because our population estimator in Eq.~\eqref{eq:Phidef} transforms correctly under unitary basis rotations.

The coherences between states $|n\rangle$ and $|m\rangle$ are linear combinations of the Hermitian operators $\hat{\sigma}^x_{nm}=|n\rangle\langle m|+|m\rangle \langle n|$ and $\hat{\sigma}^y_{nm} = -\ii\left(|n\rangle\langle m|-|m\rangle\langle n|\right)$,
\begin{subequations}
\begin{align}
    |n\rangle\langle m| &= {1\over 2}\left(\hat{\sigma}^x_{nm}+\ii\,\hat{\sigma}^y_{nm}\right)\\
    |m\rangle\langle n| &= {1\over 2}\left(\hat{\sigma}^x_{nm}-\ii\,\hat{\sigma}^y_{nm}\right),
\end{align}
\end{subequations}
and we can rewrite $\hat{\sigma}^x_{nm}$ and $\hat{\sigma}^y_{nm}$ as differences between the population operators of the rotated states $|x^{\pm}_{nm}\rangle = \sqrt{1\over 2}\left(|n\rangle\pm |m\rangle\right)$ and $|y^{\pm}_{nm}\rangle = \sqrt{1\over 2}\left(|n\rangle\pm\ii |m\rangle\right)$:
\begin{subequations}
\begin{align}
\phantom{1\over 2}
\hat{\sigma}^x_{nm} &= |x^+_{nm}\rangle\langle x^+_{nm}| - |x^-_{nm}\rangle\langle x^-_{nm}|\\
\phantom{1\over 2}
\hat{\sigma}^y_{nm} &= |\,y^+_{nm}\rangle\langle y^+_{nm}| - |\,y^-_{nm}\rangle\langle y^-_{nm}|.
\end{align}
\end{subequations}
Since these involve polulation differences, we can measure them with classical estimators in the same way as we measure populations, 
\begin{subequations} \label{eq:Sigma_def}
\begin{align}
\phantom{1\over 2}
    \hat{\sigma}^x_{nm} \mapsto \Sigma^x_{nm} = \alpha_N\left(P_{x^+_{nm}}-P_{x^-_{nm}}\right)\phantom{,}\\
\phantom{1\over 2}
    \hat{\sigma}^y_{nm} \mapsto \Sigma^y_{nm} = \alpha_N\left(P_{y^+_{nm}}-P_{y^-_{nm}}\right),
\end{align}
\end{subequations}
where $P_{x^{\pm}_{nm}}=|c_{x^{\pm}_{nm}}|^2$ and $P_{y^{\pm}_{nm}}=|c_{y^{\pm}_{nm}}|^2$ with $c_{x^{\pm}_{nm}}=\sqrt{1\over 2}(c_n\pm c_m)$ and $c_{y^{\pm}_{nm}}=\sqrt{1\over 2}(c_n\mp\ii\,c_m)$. Our mapping for the coherences is therefore 
\begin{subequations}\label{eq:nm_map}
\begin{align}
    |n\rangle\langle m| &\mapsto {1\over 2}\left(\Sigma^x_{nm}+\ii\,\Sigma^y_{nm}\right)\\
    |m\rangle\langle n| &\mapsto {1\over 2}\left(\Sigma^x_{nm}-\ii\,\Sigma^y_{nm}\right), 
\end{align}
\end{subequations}
\revb{which in view of Eq.~\eqref{eq:Sigma_def} can be written in the neater form
\begin{equation}
    |n\rangle\langle m| \mapsto \alpha_N c_n^* c_m.
\end{equation}
Together with the population estimator in Eq.~\eqref{eq:Phidef}, this leads to a general estimator for electronic observables that will give the same result in any unitarily rotated basis:
\begin{equation}
    \sum_{nm} {O}_{nm}|n\rangle\langle m| \mapsto \sum_{nm} O_{nm} \left[ \alpha_Nc_n^*c_m+\beta_N \delta_{nm}\right].
\end{equation}}

%We would stress again that this automatically treats the coherences on an equal footing to the populations: coherences are no more \lq\lq quantum mechanical" (and therefore no harder to treat classically) within this framework than populations. All that matters is that the population estimator can be applied equally well in any unitarily rotated basis, as ours can.

\subsection{Initial conditions}
To simulate a non-equilibrium process starting in a pure electronic state $|n\rangle\langle n|$, such as the bright state of a molecule that has just been photoexcited, \rev{we need to define an initial electronic distribution, $\rho_n(c)$, such that
% it is natural to sample the initial value for the electronic state vector $c$ from a distribution $\Gamma_n(c)$ such that 
\begin{equation}
    \int_{|c|=1} {{\rm d}c\over\mathcal{N}}\, \rho_n\Phi_m = \delta_{nm}.\label{eq:Gamma}
\end{equation}
In this way, the initial estimate of the population of state $m=n$ is 1, and that of all other states is zero.
%so that our initial estimate of the population of state $n$ will be 1 and that of all other states will be zero.
There are many choices of $\rho_n$ that fulfil this condition. %, of which at least two have precedents in the previous mapping literature. 
%There is an infinite number of ways to do this, at least two of which have precedents in the previous mapping literature. 
The one we will use in the following is 
\begin{equation} \label{eq:rho_Theta}
    \rho_n(c) = \Theta_n(c),
\end{equation}
corresponding to a uniform distribution in the region where $P_n$ is the largest population \revb{(see the coloured regions in Fig.~\ref{fig:simplex})}. This choice fulfils Eq.~\eqref{eq:Gamma} by the same argument as in Appendix~C, regardless of whether the initial state is a diabat or an adiabat. 
%The first is to note that the argument in Appendix~C works equally well for any set of orthonormal basis states $a$ and $b$, including the diabatic basis states $n$ and $m$. Hence Eq.~\eqref{eq:Gamma} will hold for all $n$ and $m$ if we choose to set the distribution function equal to the state projector:
% \begin{equation}
% \Gamma_n(c) = \Theta_n(c).
% \end{equation}
For a two-state problem in which the initial state is specified in the adiabatic basis, Eq.~\eqref{eq:rho_Theta} is equivalent to the prescription for $A(c)$ that Mannouch and Richardson used in Eq.~\eqref{eq:mash_observables}. However, our prescription for $B(c)$ is different from theirs, and our formulation avoids their weight function $\mathcal{W}_{AB}(c)$.}

\revb{In practice, one can sample points $c$ from $\Theta_n(c)$ is as follows. First, sample $N$ pairs $(x_k,y_k)$, where $x_k$ and $y_k$ are Gaussian deviates with zero mean and unit variance. Then set $c_k=(x_k+\ii y_k)/\sqrt{\sum_l(x_l^2+y_l^2)}$. This has $|c|=1$ and, due to the rotational invariance of the multidimensional normal distribution, is a point with uniform probability on the simplex. Finally, check which state has the largest $|c_k|^2$ -- if it is the initial state $n$, then accept the point, and if not, resample a new point in the same way until it is accepted.}

\section{Application to exciton energy transfer}

To test the multi-state algorithm introduced in Sec.~III, we have applied it to a Frenkel-exciton model of energy transfer in the Fenna--Matthews--Olson complex.\cite{ishizaki2009FMO} This has become a standard benchmark for nonadiabatic dynamics and it allows a comparison with a variety of other trajectory-based methods. The problem is challenging for conventional surface hopping due to the presence of `trivial' crossings with low hopping probabilities, which can require an extremely small time step to converge.\cite{Sindhu2022fmo} Several phase-space methods have been reported to perform well for FMO at high temperatures, including spin mapping,\cite{runeson2020,mannouch2020paperII} MMST mapping,\cite{tao2010FMO,kelly2011fmo,kim2014nH,kim2014scaling, saller2020faraday,coker2016photosynthesis} and the symmetric quasi-classical trajectory method of Cotton and Miller.\cite{cotton2016lightharvest} However, these methods struggle to recover the correct long-time equilibrium populations at low 
temperatures, and they only avoid unstable inverted potentials because all of the potential-energy surfaces in the model are harmonic with the same frequencies.

The standard FMO model is comprised of seven sites, but to facilitate a comparison with a previous surface hopping study we have also considered a reduced model with three sites. In both cases, the sites are coupled independently to identical harmonic baths. The full Hamiltonian and all model parameters are given in Appendix~\ref{app:fmo}. The initial condition for our simulations is a pure electronic state in an uncoupled classical Boltzmann bath. 
%(We did \emph{not} use a Wigner distribution for the bath because this would lead to zero-point energy leakage and artificially elevate the temperature in the long-time limit.)
The initial electronic state can be in the site or the exciton basis and we shall report results for both below. \rev{Each simulation was averaged over $10^5$ trajectories for good statistical convergence, although rough results can typically be obtained with $10^3$.}
\revb{We used a time step of 0.25 fs with the trajectory integrator and hopping protocol described in Appendix~\ref{app:details}}. Fully quantum mechanical HEOM benchmark results were computed for comparison using the \texttt{pyrho} open source software.\cite{pyrho}

\subsection{Three-state model}\label{sec:afssh}

\begin{figure}[t]
\centering
\resizebox{0.9\columnwidth}{!} {\includegraphics{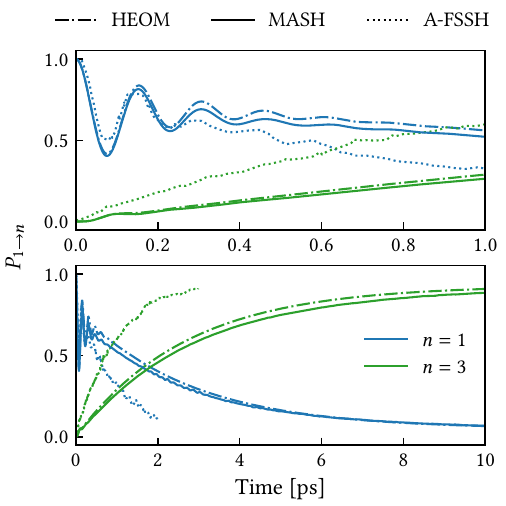}}
\caption{Comparison of HEOM (dash-dotted), MASH (solid), and A-FSSH (dotted) for population dynamics in a three-site FMO model. Results are reported in the site basis at \SI{77}{K}, starting in site 1. The A-FSSH results are from Ref.~\onlinecite{Sindhu2022fmo}.}\label{fig:fmo3}
\end{figure}

\begin{figure*}[t]
\centering
% \resizebox{0.7\columnwidth}{!} {\includegraphics{pop300Ksite.png}}~
% \resizebox{0.7\columnwidth}{!} {\includegraphics{pop300Kexc.png}}~
% \resizebox{0.7\columnwidth}{!} {\includegraphics{pop77Ksite.png}}
\resizebox{\textwidth}{!} {\includegraphics{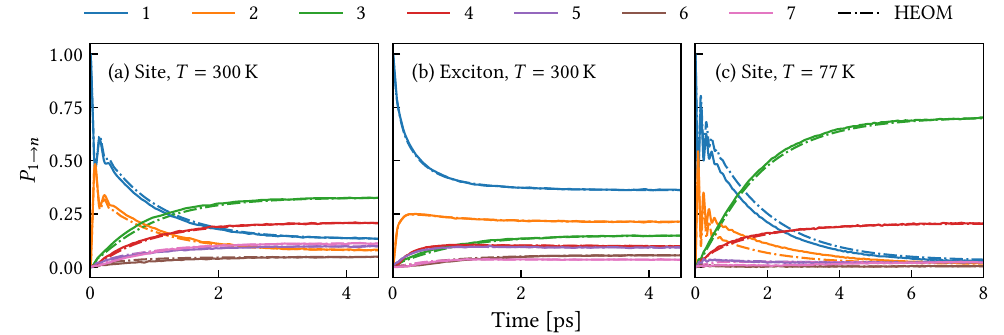}}
\caption{Comparison of MASH (solid) and HEOM (dash-dotted) for population dynamics in the seven-site FMO model at (a) \SI{300}{K} in the site basis, (b)  \SI{300}{K} in the exciton basis, and (c) \SI{77}{K} in the site basis. The initial state was site 1 for panels (a) and (c) and exciton 1 for panel (b).}\label{fig:pop7}
\end{figure*}

Sindhu and Jain have recently used a three-site FMO model to compare a variety of different decoherence corrections to fewest switches surface hopping.\cite{Sindhu2022fmo,Sindhu2023fmo} To assess the performance of MASH against FSSH, we shall use the most accurate of the methods they considered in their comparison as a reference. This method, the augmented surface hopping (A-FSSH), employs a parameter-free decoherence correction that has been found to improve upon the original FSSH algorithm for condensed-phase problems\cite{subotnik2016review} (for example, it has been found to recover Marcus theory rates in the golden-rule limit\cite{jain2015rate}). 
It should be noted that Sindhu and Jain quantized one of the nuclear modes in their FMO calculation to give a manifold of vibronic states, whereas our MASH calculations treat all nuclear modes classically. Since their quantization improved the agreement of A-FSSH with HEOM rather than harmed it,\cite{Sindhu2022fmo} we still feel this provides a fair comparison.

In Figure~\ref{fig:fmo3}, we compare MASH to the A-FSSH results from Ref.~\onlinecite{Sindhu2022fmo} as well as to HEOM. We find that MASH almost perfectly captures the initial coherent oscillations in the HEOM polulation dynamics, as well as the long-time relaxation. In contrast, A-FSSH leads to overly damped oscillations and too rapid thermalization. These results are interesting not only because they demonstrate that MASH provides a clear improvement over A-FSSH, but because it does so without decoherence corrections. This suggests that overcoherence may not be such a useful way to understand the problems of FSSH as previously thought. The issue may instead be more closely related to the inconsistency between $\langle |c_a|^2\rangle$ and the fraction of trajectories on the active surface. This is precisely the problem that MASH was designed to resolve by uniquely defining the active surface in terms of the wavefunction coefficients.\cite{Mannouch2023mash}

Another interesting observation is that, even at \SI{77}{K} where $k_\mathrm{B}T=\SI{53.5}{cm^{-1}}$ is roughly two times smaller than the characteristic phonon energy of $\hbar\omega_{\rm c}=\SI{106}{cm^{-1}}$, MASH gives accurate results despite treating the nuclei classically. This indicates that nuclear quantum effects are less important for this system than the quantization of one of the nuclear modes in Ref.~\onlinecite{Sindhu2022fmo} would suggest them to be.

\subsection{Seven-state model}

We are not aware of any calculations using conventional surface hopping for the seven-state FMO model, but it is expected that they would suffer from the same difficulties as in the three-state case. In Figure~\ref{fig:pop7}, we compare our MASH results for this model directly to those of the quantum HEOM benchmark. Panel (a) shows the population dynamics in the site basis at \SI{300}{K}, starting from site 1, and panel (b) shows the dynamics in the exciton basis at \SI{300}{K}, starting from exciton state 1. The exciton basis is defined as the eigenbasis of the system Hamiltonian in Eq.~\eqref{eq:HS}. In both cases, MASH is seen to agree almost perfectly with the fully quantum benchmark. The agreement is at least as good as or better than that obtained with previously reported mapping approaches at the same temperature.\cite{cotton2016lightharvest,coker2016photosynthesis,hanna2018decide,runeson2020,mannouch2020paperII,saller2020faraday}

\begin{figure}[b]
\centering
\resizebox{0.9\columnwidth}{!} {\includegraphics{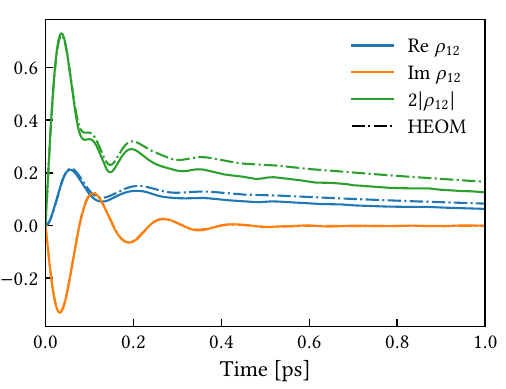}}
\caption{Comparison of MASH (solid) and HEOM (dash-dotted) results for the dynamics of the dominant coherence of the seven-state FMO model in the site basis at $\SI{300}{K}$, starting from site 1.}\label{fig:coh}
\end{figure}

Panel (c) of Figure~\ref{fig:pop7} shows the same situation as in panel (a), but at \SI{77}{K}. This low-temperature regime is more challenging for mapping approaches, several of which predict negative populations. However, MASH continues to agree well with HEOM at short times, except perhaps for the small deviation in the transfer between sites 1 and 2, and it gives the correct quantum-mechanical equilibrium populations in the long-time limit. It is worth repeating that, as in the case of the three-state model considered in Figure~\ref{fig:fmo3}, these long-time populations are obtained correctly without including nuclear quantum effects in the calculation. While nuclear quantum effects have previously been claimed to be important for the functioning of light-harvesting complexes,\cite{cao2020review,Kundu2022science}  a more recent study has shown that classical nuclei are in fact sufficient to describe realistic models of FMO at room temperature,\cite{runeson2022fmo} and our present results suggest that this conclusion extends even to cryogenic temperatures. (Including nuclear quantum effects might perhaps improve on the MASH description of the short-time population transfer between sites 1 and 2 in panel (c) of Figure~\ref{fig:pop7}, but this would only be a small correction.)

\begin{figure*}[t]
    \centering
    \includegraphics{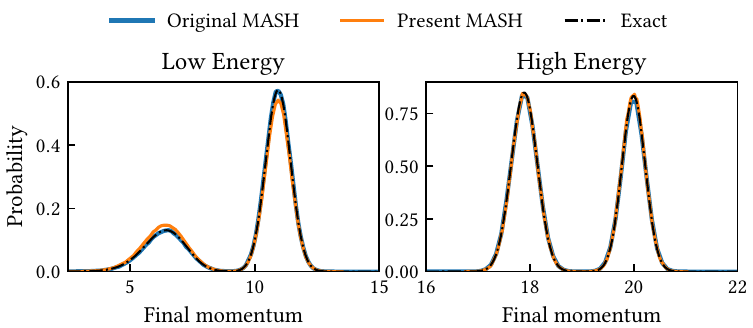}
    \caption{Final momentum distribution after a single avoided crossing in Tully's first model, for two initial energies. The original MASH results and the model parameters are from Ref.~\onlinecite{Mannouch2023mash}. \revb{These histograms were generated from $10^6$ trajectories.}}
    \label{fig:tully1}
\end{figure*}

\subsection{Coherence dynamics}

To assess the accuracy of the coherence estimator introduced in Sec.~III.D, we have used it to calculate the coherence between sites 1 and 2 of the seven-state FMO model with the initial population on site 1. To facilitate a comparison with previous spin-mapping simulations,\cite{runeson2020} the timescale of the bath was set to $\tau_{\rm c}=\SI{100}{fs}$ in these calculations, as it was in the HEOM calculations of Sarovar {\em et al.}\cite{sarovar2010nature}. Since these earlier HEOM calculations included a trapping rate that is not present in our FMO model, we recomputed the HEOM results without any trapping, and found that they differ only very slightly (by around 1\%) from the results reported in Ref.~\onlinecite{sarovar2010nature}. 

Figure~\ref{fig:coh} compares the HEOM coherence with that obtained from MASH. The agreement is not perfect, but it is clearly very good for the imaginary part of the coherence and reasonably good for the real part, especially at short times. Our coherence estimator has therefore been validated for a multi-state problem. There are no previous surface hopping results for this problem, for the same reason as there are none for the population dynamics of the seven-state model in Figure~\ref{fig:pop7}. The coherence has, however, been calculated previously using spin mapping,\cite{runeson2020} which was found to give results of similar quality to the MASH results in Figure~\ref{fig:coh}. 

\rev{\section{Application to two-level systems}
Since our scheme for measuring electronic observables differs from that of Mannouch and Richardson even for two-level systems, it is important to check the performance of our approach also in this case. For this purpose, we have applied our method to some of the two-state Tully,\cite{tully1990hopping} spin-boson, and pyrazine models considered previously by Mannouch and Richardson.\cite{Mannouch2023mash} All model definitions and parameters are the same as in Ref.~\onlinecite{Mannouch2023mash}, where further comparisons to FSSH, Ehrenfest, and spin-mapping results can be found.} \revb{Unless otherwise stated, the results were computed from $10^5$ trajectories and integrated with a timestep of \SI{0.25}{fs}.}

\rev{
\subsection{Tully models}
First, in Fig.~\ref{fig:tully1}, we consider the problem of wavepacket splitting in Tully's single avoided crossing model.\cite{tully1990hopping,miller2007scivr} Mannouch and Richardson have shown that their scheme reproduces the correct final momentum distribution at both low and high incoming kinetic energies. Our scheme uses a different trajectory weighting that reproduces their results at high energy but gives slightly less accurate peak heights at low energy. The area under each peak is proportional to the fraction of trajectories emerging on each adiabat. In the original MASH, this fraction is $\langle\Theta_a(t)\rangle$, where $t$ is a time at which the crossing is complete. In our scheme, the time-dependent populations $\langle\Phi_a(t)\rangle$ do give the correct adiabatic state branching ratio in the low energy example. However, since $\langle\Phi_a(t)\rangle$ is not the same as the fraction of trajectories emerging on each adiabat, our final momentum distribution is slightly incorrect.}

\begin{figure}[b]
    \centering
    \resizebox{0.9\columnwidth}{!}{\includegraphics{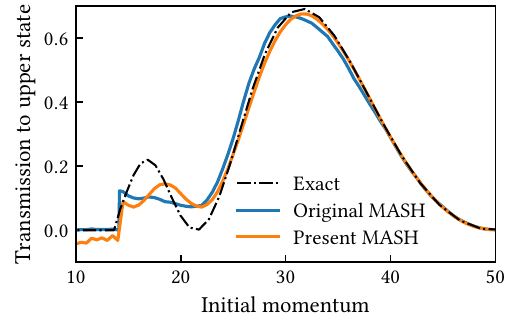}}
    \caption{Transmission probability from the lower to the upper adiabat in Tully's double avoided crossing model. The original MASH results and the model parameters are from Ref.~\onlinecite{Mannouch2023mash}.}
    \label{fig:tully2}
\end{figure}

\rev{
Fig.~\ref{fig:tully2} shows the probability of transmission on the upper adiabat for Tully's double avoided crossing model as a function of the initial momentum on the lower adiabat. Here, our results are similar to those of the original MASH method, and arguably slightly better at high momenta. For low momenta, neither method can reproduce the St\"{u}ckelberg oscillation, which is due to electronic interference. Note also that our estimator predicts negative transmission probabilities at low momenta where the upper product adiabat is energetically inaccessible, whereas Mannouch and Richardson's transmission probability goes correctly to zero.}

\rev{
Despite the issues that these tests have identified, the overall impression we get from the comparisons in Figs.~6 and~7 is that our method does not perform significantly worse for these models than the original MASH method of Mannouch and Richardson.\cite{Mannouch2023mash} Especially when one considers that these non-ergodic, one-dimensional, microcanonical models do not satisfy the assumptions we made when deriving our population estimator, and are not therefore the sort of problems for which our method was designed. The present MASH results in Fig.~6 are certainly more accurate than those of either Ehrenfest dynamics or spin mapping, for example, both of which fail to describe the wavepacket bifurcation.\cite{Mannouch2023mash}}

% \subsection{Spin-boson model}
\begin{figure*}[t]
    \centering
    \includegraphics{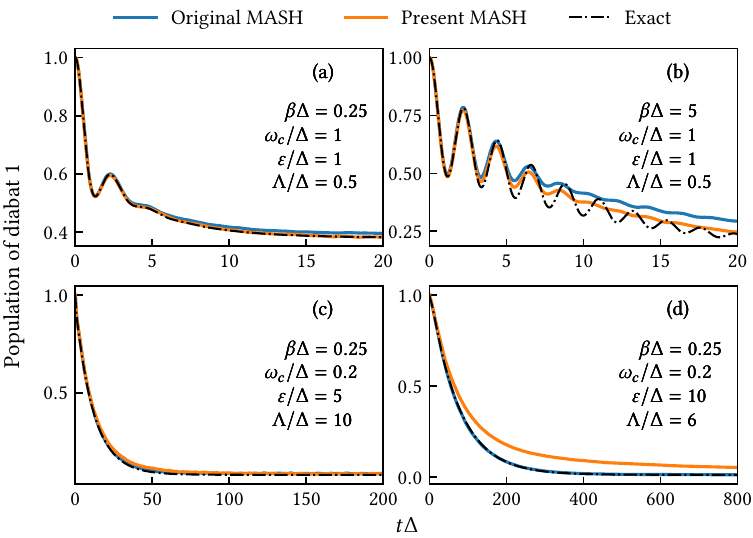}
    \caption{Population dynamics of the spin-boson model in various regimes: (a) coherent dynamics at a high temperature; (b) coherent dynamics at a low temperature; (c) \revb{activationless electron transfer; (d) electron transfer deep in the inverted regime}. The original Mash results, the exact results, and the model parameters are all from Ref.~\onlinecite{Mannouch2023mash}.}
    \label{fig:spinboson}
\end{figure*}

\begin{figure}[b]
    \centering
    \resizebox{0.9\columnwidth}{!}{\includegraphics{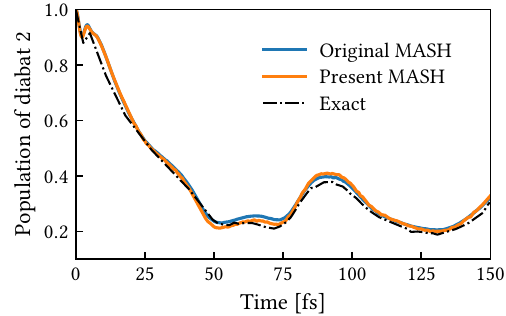}}
    \caption{Population transfer in a 24-mode pyrazine model. The original MASH results and the model parameters are from Ref.~\onlinecite{Mannouch2023mash}. The exact (MCDTH) results are from Ref.~\onlinecite{raab1999density}.}
    \label{fig:pyrazine}
\end{figure}

\rev{
\subsection{Spin-boson model}
Next, we consider the spin-boson model, for which
\begin{equation}
\hat{H}=H_0 + \Delta\,\hat{\sigma}_x + \Bigl(\varepsilon + \sum_j \xi_jq_j\Bigr)\hat{\sigma}_z,
\end{equation}
where $H_0=\frac{1}{2}\sum_j(p_j^2+\omega_j^2q_j^2)$ is a bath of oscillators with spectral density \begin{equation}
J(\omega)=\frac{\Lambda}{2}\frac{\omega\omega_{\rm c}}{\omega^2+\omega^2_{\rm c}}.
\end{equation}
For this model, we initialized the nuclei from a Wigner distribution to be consistent with Ref.~\onlinecite{Mannouch2023mash}. The resulting population dynamics is shown in Fig.~\ref{fig:spinboson} for a variety of regimes. \revb{The integration time step (in units of $\Delta^{-1}$) was 0.01 for panels (a), (b) and (d), and 0.002 for panel (c).} In the coherent examples [panels (a) and (b)], our scheme is slightly more accurate at long times than Mannouch and Richardson's. For \revb{the case of activationless electron transfer ($2\varepsilon=\Lambda)$} [panel (c)], both schemes yield similar results. However, in the Marcus inverted regime [panel (d)], our relaxation is too slow, even though it does eventually reach the correct limit (not shown). This regime is known to be challenging for trajectory-based methods and it is remarkable that Mannouch and Richardson's scheme is so accurate. The fact that the present version of MASH does not capture the inverted regime correctly is probably the most serious deficiency of the method we have found so far. It is not clear how to solve this problem without abandoning the basis-set independence of our population estimator, but it is conceivable that a quantum-jump procedure\cite{Mannouch2023mash} might improve our results.}

\rev{
\subsection{Pyrazine}
Finally, we consider ultrafast dynamics through the conical intersection in a full-dimentional (24-mode) model of pyrazine. This model includes bilinear couplings which make it challenging for traditional mapping methods. It is also typical of the sort of photochemical problems to which one might expect methods like MASH to be applied. As shown in Fig.~\ref{fig:pyrazine}, our results for this final two-state problem are of comparable quality to those of the original MASH scheme.\cite{Mannouch2023mash}
}

%\rev{
%To summarize, the accuracy of the new scheme is sometimes better and sometimes worse than in the original scheme, but usually comparable. An example of when the new scheme performs significantly worse is for the spin-boson model in the inverted regime. The new scheme however has the advantage that time-dependent observables are basis-invariant and do not require the complicated weighting procedure in Eq.~\eqref{eq:CWAB}.
%}

\section{Concluding remarks}
In this article, we have shown how to extend the MASH methodology of Mannouch and Richardson\cite{Mannouch2023mash} to general $N$-level systems. We have also proposed simplified estimators for populations and coherences that can be applied in any basis and are guaranteed to give the correct long-time equilibrium populations in this basis provided the system is ergodic and the classical approximation to the nuclear motion is valid. Our applications to FMO exciton models have shown that the resulting multi-state MASH method is at least as accurate as any previous semiclassical mapping method, and significantly more accurate than an up-to-date implementation of fewest-switches surface hopping.

Regarding the restriction to classical nuclear motion, we would point out that the majority of interesting nonadiabatic dynamics problems follow a photoexcitation step in which a vast amount of energy (significantly larger than $k_{\rm B}T$) has been deposited in the system. The resulting nuclear motion will often be sufficiently fast that the classical nuclear motion approximation is well justified, as it certainly seems to be in all of the FMO calculations we have presented here. However, nuclear quantum effects are expected to be important in some other contexts. For example, nuclear tunnelling is known to have a significant impact on the rates of electron transfer reactions in the Marcus inverted regime.\cite{Lawrence2019interpolation} It might therefore be interesting to add nuclear quantum effects to the present methodology, perhaps by adapting the ring-polymer molecular dynamics techniques that have already been developed for standard surface hopping.\cite{Shushkov2012rpmd}

Regarding how much more successful MASH is for the FMO problem than fewest switches surface hopping (as we have shown in Figure~3), we would say that Mannouch and Richardson's idea of deterministically tying the active surface to the adiabatic state with the largest population was quite inspired. This allowed them to {\em derive} the momentum rescaling and momentum reversal stages of their MASH surface hopping algortihm from first principles, to avoid the inconsistency between the stochastically averaged $\langle |c_a|^2\rangle$ and the active adiabatic surface $V_a(q)$, and thereby to eliminate the need for {\em ad hoc} `decoherence' corrections, in a single stroke. All we have done here is to show that their idea can be adapted to treat an arbitrary number of coupled electronic states in a straightforward and internally consistent way. \rev{While this has come at the cost of sacrificing some of the advantages of the original MASH scheme for two-state problems, as we have shown in Figures~6 to~8 and discussed in Sec.~V, we feel that the extension to more electronic states is worth this sacrifice because it opens up the possibility of applying the method to a far wider variety of interesting nonadiabatic problems.}

\section*{Acknowledgements}
We would like to thank Jonathan Mannouch and Jeremy Richardson for providing their data and for helpful discussions, Joseph Lawrence for pointing out the possibility of hopping between uncoupled states when $N>2$, and Oliver Riordan for providing us with the proof that $\pi_N=H_N/N$ in Appendix~C. Johan Runeson is supported by a mobility fellowship from the Swiss National Science Foundation.

\section*{Author declarations}

\subsection*{Conflict of interest}

The authors have no conflicts to disclose.

\section*{Data availability}

The data that support the findings of this study are available within the article.
\revb{A source code containing an implementation of the present algorithm and relevant examples is publicly available at the Github repository \texttt{github.com/jruneson/multimash}.}

\appendix
\section{Integrals over quantum states}

Our integrals over electronic states are integrals over the real and imaginary parts of each $c_n=x_n+\ii y_n$ subject to the constraint that $|c|=1$. They can be written in unnormalised form as 
\begin{equation}
\int_{|c|=1} \rd c\,f = \int \prod_n (\rd x_n \rd y_n) \,\delta\left(r-1\right)\,f,
\end{equation}
where $r=[\,\sum_n (x_n^2+y_n^2)\,]^{1/2}$ and $f=f(c)$. The transformation $x_n=P_n^{1/2}\cos\phi_n$ and $y_n=P_n^{1/2}\sin\phi_n$ 
converts this to an integral over populations and phases,
\begin{equation}
\int_{|c|=1} \rd c\,f = 2 \int_{\sum_n P_n=1} \prod_n \left(\frac{1}{2}\rd P_n \rd\phi_n \right)\,f,
\end{equation}
where the first factor of 2 comes from squaring the constraint, $\int \rd r\, \delta(r-1) = 2\int \rd r\, \delta(r^2-1)$, and the factors of 1/2 come from the Jacobian of the transformation. When the integrand is purely a function of the populations, $f=f(P)$, the phases can be integrated out to leave 
\begin{equation}
\int_{|c|=1}\rd c\,f  = 2\pi^N \int_{\sum_n P_n=1} \rd P\,f,
\end{equation}
and when $f=1$ the population integral gives the volume $1/(N-1)!$ of the simplex in Eq.~(\ref{eq:Simplex}):
\begin{equation}
\int_{|c|=1}\rd c = {2\pi^N\over (N-1)!}.
\end{equation}
To ensure the correct trace of the identity operator on the electronic space, we choose to normalise the integrals such that
\begin{equation}
\int_{|c|=1} {\rd c\over\mathcal{N}} = N!\int_{\sum_n P_n=1} \rd P = N,
\end{equation}
which gives $\mathcal{N}=2\pi^N/N!$. 

\section{An alternative constraint}

The goal here is to show that Eq.~\eqref{eq:constraint1} will be satisfied for all $a$ and $n$ provided Eq.~\eqref{eq:constraint2} is satisfied for all $a$ and $b$. From the definition of $\Phi_n$ as $\alpha_NP_n+\beta_N$, and the fact that $\int_{|c|=1} {\rd c/\mathcal{N}} \,\Theta_a=1$, we can start by writing
\begin{equation}
    \int_{|c|=1} {\rd c\over\mathcal{N}}\,\Theta_a\Phi_n =
    \beta_N+\alpha_N\int_{|c|=1} {\rd c\over\mathcal{N}}\,\Theta_a P_n.
\end{equation}
Now transforming $P_n$ into the adiabatic representation gives
\begin{equation}
    P_n = |c_n|^2 = \sum_{bb'} c_{b}^*\langle b(q)|n\rangle\langle n|b'(q)\rangle c_{b'}.
\end{equation}
When this is substituted into the integral on the right hand side of Eq.~(B1), only the diagonal terms (with $b'=b$ and $c_{b}^*c_{b'}=|c_{b}|^2=P_{b}$) survive, because the phase factors in the off-diagonal terms integrate to zero. Hence
\begin{equation}
\int_{|c|=1} {\rd c\over\mathcal{N}}\,\Theta_aP_n
= \sum_{b}|\langle b(q)|n\rangle |^2\int_{|c|=1} {\rd c\over\mathcal{N}}\,\Theta_aP_b.
\end{equation}
Substituting this back into Eq.~(B1) and rearranging, using the definition of $\Phi_b$ as $\alpha_NP_b+\beta_N$ and the fact that $\sum_b |\langle b(q)|n\rangle |^2=1$, gives
\begin{equation}
\int_{|c|=1} {\rd c\over\mathcal{N}}\,\Theta_a\Phi_n
= \sum_{b}|\langle b(q)|n\rangle |^2\int_{|c|=1} {\rd c\over\mathcal{N}}\,\Theta_a\Phi_b.
\end{equation}
If Eq.~\eqref{eq:constraint2} is satisfied, this reduces to $|\langle a(q)|n\rangle |^2$, which completes what we set out to show.

\section{Derivation of the population estimator}

We now turn to the problem of finding values for $\alpha_N$ and $\beta_N$ such that the constraints in Eq.~(\ref{eq:constraint2}) are satisfied for all $a$ and $b$. This is possible because we can rewrite all $N^2$ constraints in terms of the single parameter
\begin{equation}
\int_{|c|=1} {{\rm d}c\over\mathcal{N}}\, \Theta_aP_a=\pi_N,
\end{equation}
which is the same for all $a$ by symmetry. Indeed, since $\sum_{b=1}^N P_b=1$ we have that
\begin{equation}
\int_{|c|=1} {{\rm d}c\over\mathcal{N}}\, \Theta_aP_b = %\pi_N\delta_{ab}-{\pi_N-1\over N-1}(1-\delta_{nm})
{N\pi_N-1\over N-1}\,\delta_{ab}-{\pi_N-1\over N-1},
\end{equation}
and since $\int_{|c|=1} {\rd c/\mathcal{N}}\,\Theta_a=1$ we have that
\begin{align}
\int_{|c|=1} {{\rm d}c\over \mathcal{N}}\, \Theta_a\Phi_b &= \alpha_N\left({N\pi_N-1\over N-1}\right)\delta_{ab}\nonumber\\
&-\alpha_N\left({\pi_N-1\over N-1}\right)+\beta_N,
\end{align}
which reduces to $\delta_{ab}$ when we set
\begin{equation}
    \alpha_N = {N-1\over N\pi_N-1}\quad\hbox{and}\quad \beta_N = {\pi_N-1\over N\pi_N-1}.
\end{equation}

It remains to calculate $\pi_N$, which can be interpreted in view of its definition in Eq.~(C1) as the expectation value of the largest coordinate on the simplex in Eq.~(\ref{eq:Simplex}). This is the same (by symmetry) as the average value of $P_1$ in the region where $P_1\ge P_2\ge \ldots \ge P_N\ge 0$ and $\sum_{n=1}^N P_n=1$, which can found as follows. Let $P_{N+1}=0$ and consider the average value of $\delta P_n=P_n-P_{n+1}$ in the region where $\delta P_n\ge 0$ and $\sum_{n=1}^N n\,\delta P_n=1$. This is just a shrunken simplex with vertices at $(0,\ldots,0,1/n,0,\ldots,0)$, where $1/n$ is in the $n$-th position. So we can stretch each coordinate $\delta P_n$ by a factor of $n$ to recover a standard simplex. The average value of any coordinate in a standard simplex is $1/N$, so the average value of $\delta P_n$ in the shrunken simplex is $1/(nN)$, and since $P_1=\sum_{n=1}^N \delta P_n$ this gives
\begin{equation}
\pi_N = %\sum_{n=1}^N {1\over nN} \equiv 
{H_N\over N},
\end{equation}
where $H_N=\sum_{n=1}^N 1/n$ is the $N$-th harmonic number. 

\section{FMO model}\label{app:fmo}

The FMO model is a standard Frenkel-exciton Hamiltonian in the single-excitation manifold,
\begin{equation}
\hat{H} = \hat{H}_{\rm s} + \hat{H}_{\rm b} + \hat{H}_{\rm sb}.
\end{equation}
The seven-site system Hamiltonian $\hat{H}_{\rm s}$ is given in the site basis in units of cm$^{-1}$ as the $7\times 7$ matrix\cite{adolphs2006fmo}
\begin{equation}\label{eq:HS}
    %\hat{H}_\text{s} = 
    \begin{pmatrix} 
    200 & -87.7 & 5.5 & -5.9 & 6.7 & -13.7 & -9.9 \\ 
    -87.7 & 320 & 30.8 & 8.2 & 0.7 & 11.8 & 4.3 \\
    5.5 & 30.8 & 0 & -53.5 & -2.2 & -9.6 & 6.0 \\
    -5.9 & 8.2 & -53.5 & 110 & -70.7 & -17.0 & -63.3 \\
    6.7 & 0.7 & -2.2 & -70.7 & 270 & 81.1 & -1.3 \\
    -13.7 & 11.8 & -9.6 & -17.0 & 81.1 & 420 & 39.7 \\
    -9.9 & 4.3 & 6.0 & -63.3 & -1.3 & 39.7 & 230
    \end{pmatrix},
\end{equation}
where we have subtracted the lowest bacteriochlorophyll excitation energy of $\SI{12210}{cm^{-1}}$ from the diagonal entries to save paper. The three-site model uses the upper left $3\times 3$ submatrix.

The bath Hamiltonian is
\begin{equation}
    \hat{H}_\text{b} = \sum_{n=1}^N \sum_{j=1}^f  \left( \frac{1}{2} p_{j,n}^2  + \frac{1}{2}\omega_j^2 q_{j,n}^2\right),
\end{equation}
and the system--bath coupling is
\begin{equation}
    \hat{H}_{\rm sb} = \sum_{n=1}^N \sum_{j=1}^f \kappa_{j}q_{j,n}|n\rangle\langle n|,
\end{equation}
where $|n\rangle\langle n|$ is a projection onto site $n$. 
The frequencies $\omega_j$ and the coupling coefficients $\kappa_{j}$ of each bath are determined by the same spectral density
\begin{equation}
J(\omega) = \frac{\pi}{2}\sum_j \frac{\kappa_j^2}{\omega_j}\delta(\omega-\omega_j),
\end{equation}
which is taken to be of Debye form
\begin{equation}
J(\omega) = 2\lambda \frac{\omega\omega_{\rm c}}{\omega^2+\omega^2_{\rm c}}
\end{equation}
with a reorganization energy of $\lambda=\SI{35}{cm^{-1}}$ and a characteristic phonon frequency of $\omega_{\rm c}=1/\tau_{\rm c}$. Figures~\ref{fig:fmo3} and~\ref{fig:pop7} used $\tau_{\rm c}=\SI{50}{fs}$ ($\omega_{\rm c}=\SI{106.14}{cm^{-1}}$), whereas Figure~\ref{fig:coh} used $\tau_{\rm c}=\SI{100}{fs}$ ($\omega_{\rm c}=\SI{53.07}{cm^{-1}})$ for reasons discussed in the text. The Debye bath was discretized into 60 modes per site with the same procedure as described \rev{on p.~51} in Ref.~\onlinecite{Hele2013masters}.

\section{Implementation details}\label{app:details}
To integrate the MASH equations of motion for a finite time step $\Delta t$, we used a simple velocity-Verlet scheme
\begin{subequations}\label{eq:Verlet}
\begin{align}
    c_n &\leftarrow \sum_m [\e^{-\ii \hat{V}(q)\Delta t/2}]_{nm}c_m  \\
    p &\leftarrow p - \nabla V_a(q)\Delta t/2 \\
    q &\leftarrow q + (p/m)\Delta t\\
    p &\leftarrow p - \nabla V_a(q)\Delta t/2 \\
    c_n &\leftarrow \sum_m [\e^{-\ii \hat{V}(q)\Delta t/2}]_{nm}c_m  
\end{align}
\end{subequations}
where $a$ is the index of the active surface.
 
\revb{After each time step, the adiabatic populations were calculated, and if a new state $b$ had reached a higher population than state $a$, we used 10 bisections  to find the crossing point $\delta t<\Delta t$ where $P_a(\delta t)=P_b(\delta t)$. In each bisection iteration, Eq.~\eqref{eq:Verlet} was used to propagate the system on state $a$ from the original starting point, but through a time step of $\delta t$ rather than $\Delta t$. %In practice, we found that 10 bisection iterations were enough to locate the crossing point sufficiently well to implement the hop and obtain a stable integration of the MASH equations of motion. 
After the hop, the remainder of the original time step, $\Delta t-\delta t$, was processed in the same way, starting on the new active surface if the hop had been successful. Occasionally, after an unsuccessful hop, we found that the trajectory would attempt to hop again multiple times before reaching the end of the full time step $\Delta t$. There may be better ways to deal with this, but in the present calculations we simply abandoned the trajectory after 30 unsuccessful hopping attempts within any given time step. This happened for less than 1.6\% of the trajectories in the worst case.}

The hops were implemented by switching the active surface and rescaling the momentum to conserve the total energy, or reversing the momentum and abandoning the hop if the kinetic energy was insufficient to cross the potential step. The time derivative of the adiabatic population difference associated with a hop is
\begin{equation}\label{eq:Pdot_multilevel}
\dot{P}_a - \dot{P}_b = 
2 \sum_j \frac{p_j}{m_j} \sum_{a'}\Re\left[c_{a'}^*d^j_{a'a}c_a-c_{a'}^*d^j_{a'b}c_b\right],
\end{equation}
\rev{from which one can identify
the component of the momentum that needs to be rescaled or reversed. In practice, we project the mass-scaled momenta $\tilde{p}_j=p_j/\sqrt{m_j}$ onto the direction $\hat{\delta}_{ab}=\delta_{ab}/|\delta_{ab}|$  of a vector with elements
\begin{equation}
\delta_{ab}^j = {1\over \sqrt{m_j}}\sum_{a'} \Re\left[c_{a'}^*d^j_{a'a}c_a-c_{a'}^*d^j_{a'b}c_b\right],
\end{equation}
and rescale the magnitude of this projection from $|\tilde{p}_{\rm old}|$ to  $|\tilde{p}_{\rm new}| = \sqrt{|\tilde{p}_{\rm old}|^2 + 2(V_a-V_b)}$, leaving its orthogonal complement unchanged. If the argument of the square root is negative, the hop is abandoned and the projection is instead reversed.} The resulting rescaling/reversal can be regarded as arising from a classical particle incident on a step barrier, as in the two-level case.\cite{Mannouch2023mash}
\rev{This way of treating momentum reversal arises naturally from the MASH equations of motion and it differs from the conventional protocols used in FSSH.\cite{Jasper2003momentum}}

\bibliography{runerefs}

%\begin{thebibliography}{99}
%\end{thebibliography}

\end{document}